
\input phyzzx
\hoffset=0.2truein
\voffset=0.1truein
\hsize=6truein
\def\TITLEPAGE{\frontpagetrue}

\def\cinve#1{\hbox to\hsize{\tenpoint \baselineskip=12pt
	   \hfil\vtop{\hbox{\strut CIEA-FIS/98-#1}
	   \hbox{\strut CPHT-S6110598}
           \hbox{\strut May 1998}}}}

\def\cepth#1{\hbox to\hsize{\tenpoint \baselineskip=12pt
        \hfil\vtop{\hbox{\strut CPHT-98-#1}
        \hbox{\strut Research Report}
        \hbox{\strut March 1997}}}}

\def\CPHT{\smallskip
        \address{$(\bullet)$ Centre de Physique Th\'eorique, Ecole
         Polytechnique,\break
         91128 Palaiseau, France}}

\def\CIEA{\smallskip
	 \address{$(\star)$ Departamento de F{\'\i}sica,\break 
          Centro de Investigaci\'on y Estudios Avanzados del IPN,\break
Apdo. Postal 14-740, c.p. 07000 M\'exico, D.F., MEXICO}}

\def\AUTHOR#1{\vskip .5in \centerline{#1}}

\def\ABSTRACT#1{\vskip .5in \vfil \centerline{\twelvepoint \bf Abstract}
	#1 \vfil}
\def\ENDTITLEPAGE{\vfil\eject\pageno=1}

\def\sqr#1#2{{\vcenter{\hrule height.#2pt
      \hbox{\vrule width.#2pt height#1pt \kern#1pt
        \vrule width.#2pt}
      \hrule height.#2pt}}}

\def\section#1#2{
\noindent\hbox{\hbox{\bf #1}\hskip 10pt\vtop{\hsize=5in
\baselineskip=12pt \noindent \bf #2 \hfil}\hfil}
\medskip}

\def\underwig#1{	
	\setbox0=\hbox{\rm \strut}
	\hbox to 0pt{$#1$\hss} \lower \ht0 \hbox{\rm \char'176}}

\def\bunderwig#1{	
	\setbox0=\hbox{\rm \strut}
	\hbox to 1.5pt{$#1$\hss} \lower 12.8pt
	 \hbox{\seventeenrm \char'176}\hbox to 2pt{\hfil}}


\def\PRL#1#2#3{{\it Phys. Rev. Lett.} {\bf#1} (#2) #3}

\def\CQG#1#2#3{{\it Class. Quan. Grav.}{\bf #1} (#2) #3}

\def\JMP#1#2#3{{\it J. Math. Phys.} {\bf #1} (#2) #3}


\def\thep{1 + \half \zeta\bar\zeta}
\def\thepc{1 - \half \zeta\bar\zeta}
\def\theq{1 + ({\zeta\bar\zeta\over 2})^2}
\def\theppc{1 - ({\zeta\bar\zeta\over 2})^2}
\def\Re{{\rm Re}\;}
\def\Im{{\rm Im}\;}
\def\by{{\zeta}}
\def\bby{{\overline{\zeta}}}
\def\bd{{\overline D}}
\def\bl{{\overline L}}
\def\wy{{\scriptscriptstyle \zeta}}
\def\bry{{{\overline{\wy}}}}

\def\bL{{\overline L}}
\def\sym{{\mathop{\otimes}\limits_{s}}\,}
\def\p{\partial}
\def\h{\textstyle {1 \over 2}}

\REF\Ha{I. Hauser, \PRL{33}{1974}{1112}, \JMP {19}{1978} {661}.}

\REF\Mc{McIntosh, C.B.G., \CQG{2}{1985}{87-97}.}

\REF\St{H. Stephani, \CQG{10}{1993}{2187-2190}.}

\REF\DKS{G. C. Debney, R. P. Kerr and A. Schild, \JMP{10}
{1969}{1842-1854}.}

\REF\KSMH{D. Kramer, H. Stephani, M. MacCallum and E. Herlt, {\it Exact
solutions of Einstein's field equations}, (VEB Deutscher Verlag der
Wissenschaften, Berlin; Cambridge Univ. Press, Cambridge, 1980). }

\REF\FPP{J. D. Finley, J. F. Pleba\'nski and Maciej Przanowski, {\it 
Approach to twisting and diverging, type N, vacuum 
Einstein equations: a (third-order) resolution of Stephani's `paradox'}, 
\CQG{14}{1997}{489-498}, gr-qc/9605054.}

\REF\BIPR{J. Bi\v c\'ak and V. Pravda, {\it Curvature invariants in
type-N spacetimes}, \CQG{15}{1998}{1539-1555}, gr-qc/9804005.}


\TITLEPAGE
\cinve{30}
\bigskip 
\titlestyle {A note on the iterative approach to twisting 
type N solutions} 

\AUTHOR{
L. Palacios${}^\star$
\footnote\bullet{Present address\hfil\break
palacios@pth.polytechnique.fr} 
and 
J.F. Pleba\'nski
\footnote\star{pleban@fis.cinvestav.mx}} 
\CIEA 

\CPHT

\ABSTRACT{We present the results of the 
computation of a twisting 
type N solution to vacuum Einstein equations following an iterative
approach. Our results show that the higher order terms fail to
provide a full exact  solution with non-vanishing twist. Nevertheless,
our fourth-order solution still represents a regular and twisting type N
solution.}

\medskip
\ENDTITLEPAGE

\eject

\medskip
\noindent {\bf 1. Introduction }

There is an interesting unsolved problem of finding solutions of vacuum Einstein 
equations that represent gravitational radiation outside a bounded source. 
The non-twisting solutions of type N, either approximate or exact,  
have always singularities in space at infinity, hence they are not good 
candidates for this purpose. 
Also, there have been various different efforts to find
twisting type N 
solutions [\Ha, \Mc] and, so far, there is only one exact solution with 
non-vanishing twist reported by Hauser [\Ha]. 
Hauser's solution also has singularities which extend to spatial
infinity. Thus, it fails to provide an exact description of the
gravitational radiation.

Recently, Stephani [\St]  studied a linearised form of twisting type N 
solutions and  claimed that either there will always exist singular lines in 
space for 
twisting solutions or that twisting type N fields do not describe a 
radiation field outside a bounded source. Later on, in [\FPP] an iterative 
approach has been 
proposed in an attempt to solve this puzzle.  

Using this iterative approach, 
the linear solution in 
[\St] has been reproduced and higher order terms of 
this iteration process have been described in [\FPP]. In particular,
the authors have
used a 
first-order 
solution, flat and non-twisting, to derive an approximate non-flat
twisting solution. This solution is given by   
regular functions for sufficiently large spheres
(i.e. spheres defined by the ($\zeta, \overline{\zeta}$) coordinates for
large
values of the affine parameter $r$). This is an
interesting example of twisting solutions, where behaviour is contrary to 
the expectations generated by 
Stephani's paper.
  
After all this, one is wondering whether this iterative approach could
shed light 
on the general behaviour of solutions describing gravitational radiation from 
a bounded source. In this note, we present the solution obtained by
iterating the third-order 
solution in [\FPP] up to higher orders.  
Unfortunately, nothing can be said about the existence of a regular exact
solution with a non-vanishing twist parameter.

\medskip
\noindent {\bf 2. The vacuum field equations}
\medskip
 Following the notation in [\FPP], we write the
metric $\bf g$ in terms of a complex null tetrad, $e^\mu$ as follows:
$${\bf g} = g_{\mu\nu}e^\mu\sym e^\nu = 2e^1\sym e^2 + 2e^3\sym e^4\qquad,
\qquad\overline{e^1}
= e^2\,,\; \hbox{and } e^3,\ e^4\  \hbox {~real,}\eqn\metri$$
the overbar indicates complex conjugation.  For
any vacuum, type-N space-time with
non-vanishing complex expansion, i.e., where $\, Z \equiv -\Gamma_{421}\ne 0$,
the null tetrad $e^\mu$  can always be written as [\DKS]
$$\eqalign{e^1  = {1\over PZ}d\zeta\quad,\quad e^2 = & {1\over 
P\overline{Z}}d\bby,\quad - e^3 = du + L\,d\zeta + \bL\, 
d\overline{\zeta}\quad,\cr
e^4 & = dr + W \,d\zeta +\overline{W}\,d\overline{\zeta} - H\,e^3\quad,\cr}
\eqn\tetrad$$
where the metric functions are given by
$$ \matrix{Z^{-1} = (r - i\Sigma)\,, & 2i\Sigma = P^2(\bd L - D\bl) \cr\cr
W = -{\overline{Z}}^{-1}L_{,u} + iD\Sigma\,, & D\equiv \p_\wy - L\,\p_u \cr\cr
H = -r\p_u(\ln P) + \h K\,, & K = 2P^2\,\hbox{Re}[D(\bd \ln P - \bl_{,u})]\cr}
\eqn\struc$$
where the subscripts indicate partial differentiation.  Within this tetrad,
setting $P\equiv V_{,u}$,  the
remaining Einstein vacuum field equations take the form 
$$\eqalign{\bd\Big\{P^{-1} & \p_uDDV\Big\} {}= 0\quad,\cr
\noalign{\vskip-3truept}
\bd\,\bd D D V &{}= DD\bd\,\bd V\quad.\cr}\eqn\eineqns$$
The curvature and twist are given below and we must require not to be zero
$$\eqalign{ \h\overline{C^{(1)}}\  \ \equiv \Psi_4 & {} =  -\overline{Z}P^2
\p_u\p_u\left\{P^{-1}\bd\,\bd
V\right\} \ne 0\quad,\cr
2i\,\Sigma =  & P^2\left(\bd L - D\bL\right) \ne 0\;.\cr}\eqn\inva$$
The particular form of this (non-zero) curvature component is however not an 
invariant of this problem, since it may be changed, remaining non-zero,
by a choice of transformation to new (allowed) coordinates.
It is known that
for type N spacetimes there are no non-zero invariants of the
curvature tensor, which requires us to use the  
gauge invariants of the problem.
Therefore, we should use 
the two (gauge) invariants of this problem, the ratios
$\Sigma/r$ and $K/r^2$ [\KSMH] with the condition that 
$\Sigma$ and $K$  be regular functions of their arguments, (see also
[\BIPR]).

Fixing the gauge such that 
$$ P = \thep; \qquad V = P u\eqn\pgauged$$
and defining 
$$ \Phi\ \equiv L,_\wy + {\bby\over 1+\h\by\bby}L - L\p_u L\eqn\laphi $$
then the remaining Einstein vacuum field equations \eineqns\ can be written 
as (see [\FPP] for details)
$$\p_\bry\p_u\Phi - \bl\p_u^{\,2}\Phi = 0\quad.\eqn\eea$$
$$ \hbox{Im}\left\{\p_{\bry}\p_\bry\Phi + {\by\over 1+\h\by\bby}\p_\bry
\Phi - \overline{\Phi}\p_u\Phi - \bl \p_\bry\p_u\Phi\right\} = 0\;.
\eqn\eeb$$
The N-th order approximation will then be one that approximates the 
complete solution, up to that order of iteration, as a sum of all the 
preceding orders
$$\Phi \approx \sum_{j=1}^N\,\Phi^{(j)} \;,\quad L \approx 
\sum_{j=1}^N\,L^{(j)}\eqn\thesums$$
where $L^{(0)} = 0$ and $\Phi^{(0)} = 0$ give the flat space-time. The case 
$N = 1$ accounts for the linearised version in [\St] where $\Phi^{(1)}$
and $L^{(1)}$ can be obtained 
from the equations 
$$ \p_\bry\p_u\Phi^{(1)} = 0 \qquad
\hbox{~and~~}\qquad P^{-2}\left\{P^2\,L^{(1)}\right\}_{,\wy} = 
\Phi^{(1)}\;.\eqn\orderone$$
At the n-th level of iteration, $n \geq 2$ we have
$$\eqalign{\p_\bry\p_u\Phi^{(n)} =\  & \sum_{j=1}^{n-1}\,\bl^{(n-j)}
\p_u^{\,2} \Phi^{(j)}\;,\cr
P^{-2}\left\{P^2\,L^{(n)}\right\}_{,\wy} = \p_\wy L^{(n)}  & {}+ 
{\bby\overwithdelims()1+\h\by\bby}\,L^{(n)} = \ \Phi^{(n)}
 + \sum_{j=1}^{n-1} L^{(n-j)}\p_u L^{(j)}\;.\cr}\eqn\ordern$$
whose solutions are obtained from straightforward calculations up to some 
arbitrary functions that can be fixed with the eqn. \eeb\ constraint.
Using \inva, the curvature may be explicitly
re-written in terms of the current variables:
$$r\Psi_4 =  P^2\left(1+i\Sigma/r\right)^{-
1}\p_u^{\,2}\overline\Phi\;.\eqn\curvature$$
the n-th level of the gauge invariants $\Sigma = \sum\Sigma^{(n)}$ and $K = 
\sum K^{(n)}$ at the lowest order terms are $\Sigma^{(0)} = 0 $ and 
$K^{(0)} = 1$. The higher order terms are given by
$$ \Sigma^{(1)} = P^2 \Im [\p_{\bar\zeta} L^{(1)}]\eqn\thesone$$
$$K^{(1)} = -2 P^2 \Re [\p_\by\p_u \bL^{(1)}]\eqn\thekone$$  
and for $n \geq 2$
$$ \Sigma^{(n)} = P^2 \Big\{\Im [\p_{\bar\zeta} L^{(n)}] - \sum_{j =1}^{n-1} 
\Im [\bL^{(n-j)}\p_u L^{(j)}]\Big\}\eqn\thesigman$$
$$K^{(n)} = -2 P^2 \Big\{\Re [\p_\by\p_u \bL^{(n)}] -  \sum_{j =1}^{n-1}   
\Re [L^{(n-j)}\p^2_u \bL^{(j)}]\Big\}\eqn\thekn$$
Next, given a solution at first order one could in principle compute a
series 
that may converge or not.

\vfill\eject
\noindent {\bf 3.  Fourth-order solution}
\medskip
Starting with the ansatz, 
$$L^{(1)} = {1\over\zeta}\qquad \hbox{and} \qquad \Phi^{(1)} = 
{a_1\over\zeta}({\bar\zeta\over\thep})\eqn\ansa$$
and following the iterative approach described above, the authors of
[\FPP]
computed a twisting solution to the third level of iteration. 
Now, using \ordern\ 
and solving for $L^{(4)}$ and $\Phi^{(4)}$ we have 
the fourth-order terms given as
$$\eqalign{ L^{(4)} &= {a^{(4)}\over\zeta} + 
{\bar\zeta\over(\thep)^2} f^{(4)}(u) + {\thepc\over\thep}( 
{a^{(2)}\over\zeta} {d f^{(2)}\over du} + {a^{(1)}\over\zeta} {d
f^{(3)}\over 
du})\cr &- {\bar\zeta\over(\thep)^3} (2 f^{(2)}(u){d f^{(2)}\over du}) - 
{\theq\over\zeta(\thep)^2} a^{(1)} {d^2 f^{(2)}\over du^2}(a^{(1)} -  
\overline{a^{(1)}})\cr
&+ {a^{(1)} \overline{a^{(1)}}\over\zeta(\thep)^2}{d^2 f^{(2)}\over du^2} 
[(\theppc) \ln(\zeta\bar\zeta) - {\zeta\bar\zeta\over 
2}(\ln(\zeta\bar\zeta))^2]\cr}\eqn\elcua$$

\smallskip

$$\Phi^{(4)} = -{a^{(2)}\over\zeta^2}{d f^{(2)}\over du} -
{a^{(1)}\over\zeta^2}{d f^{(3)}\over du} +
{a^{(4)}\over\zeta}({\bar\zeta\over\thep} - {1\over\zeta})
- {a^{(1)} \overline{a^{(1)}}\over\zeta^2}{d^2 f^{(2)}\over du^2}
\ln(\zeta\bar\zeta)\eqn\ficua $$
that satisfies the constraint \eeb.
Now, one can compute $\Sigma^{(4)}$ and $ K^{(4)}$ from \thesigman\ and 
\thekn, thus

$$\eqalign{ \Sigma^{(4)} &= {(\thepc)\over (\thep)} \Im ( f^{(4)}(u)) - 2 (\Re 
a^{(1)})(\Im {d f^{(3)}\over du}) - 2 (\Re a^{(2)})(\Im {d f^{(2)}\over du})\cr
&- {(1 - \zeta\bar\zeta)\over(\thep)^2} \Im({d (f^{(2)})^2\over du})+ 2 
{\thepc\over\thep} \Re (a^{(1)} {d^2 f^{(2)}\over du^2})(\Im a^{(1)}) \cr
&- {\zeta\bar\zeta\over(\thep)^2}\Im (\overline{f^{(2)}} {d f^{(2)}\over du})-
a^{(1)} \overline{a^{(1)}} \Im({d^2 f^{(2)}\over du^2})[2 \ln(\zeta\bar\zeta) 
+ \half {\thepc\over \thep}(\ln(\zeta\bar\zeta))^2]\cr}\eqn\sigcua $$
\smallskip 

$$\eqalign{ K^{(4)} &= - 2 {(\thepc)\over (\thep)} \Re ( {d f^{(4)}\over du}) +
4 (\Re a^{(2)})(\Re {d^2 f^{(2)}\over du^2}) + 4 (\Re a^{(1)})(\Re {d^2 
f^{(3)}\over du^2}) \cr
&+ 2 {(1 - \zeta\bar\zeta)\over(\thep)^2} \Re({d^2 (f^{(2)})^2\over du^2}) +
{2 \zeta\bar\zeta\over(\thep)^2}\Re(f^{(2)}{d^2\overline{f^{(2)}}\over 
du^2}) \cr
&+ 4 {(\thepc)\over (\thep)} \Re (i\overline{a^{(1)}} {d^3\overline{f^{(2)}}\over
du^3 }) \cr &+ 2 a^{(1)} \overline{a^{(1)}} \Re({d^3 f^{(2)}\over du^3})[2 
\ln(\zeta\bar\zeta) + \half {\thepc\over
\thep}(\ln(\zeta\bar\zeta))^2]\cr}\eqn\kacua$$
 
Next, adding all the preceding orders of interation we have the fourth-order 
approximation as follows
$$\eqalign{ L &\approx {a_1\over\zeta} + {\bar\zeta\over(\thep)^2} 
f_2(u) + {\thepc\over\thep}
({a_1\over\zeta} {d f_2\over du}) \cr 
&- {\bar\zeta\over(\thep)^3} (2 f^{(2)}(u){d f^{(2)}\over du}) - 
{\theq\over\zeta(\thep)^2} a^{(1)} {d^2 f^{(2)}\over du^2}(a^{(1)} -  
\overline{a^{(1)}})\cr
&+ {a^{(1)} \overline{a^{(1)}}\over\zeta(\thep)^2}{d^2 f^{(2)}\over du^2} 
[(\theppc) \ln(\zeta\bar\zeta) - {\zeta\bar\zeta\over 
2}(\ln(\zeta\bar\zeta))^2]\cr}\eqn\elto $$

\smallskip

$$\Phi \approx -{a_1\over\zeta^2}{d f_2\over du} +
{a_1\over\zeta}({\bar\zeta\over\thep} - {1\over\zeta})
- {a^{(1)} \overline{a^{(1)}}\over\zeta^2}{d^2 f^{(2)}\over du^2}
\ln(\zeta\bar\zeta)\eqn\sigto $$

\smallskip
 
$$\eqalign{ \Sigma &\approx {(\thepc)\over (\thep)} \Im ( f_2(u)) - 2 (\Re
a_1)(\Im {d f_2(u)\over du}) \cr
&- {(1 - \zeta\bar\zeta)\over(\thep)^2} \Im({d (f^{(2)})^2\over du})+ 2
{\thepc\over\thep} \Re (a^{(1)} {d^2 f^{(2)}\over du^2})(\Im a^{(1)}) \cr
&- {\zeta\bar\zeta\over(\thep)^2}\Im (\overline{f^{(2)}} {d f^{(2)}\over du})-
a^{(1)} \overline{a^{(1)}} \Im({d^2 f^{(2)}\over du^2})[2 \ln(\zeta\bar\zeta)
+ \half {\thepc\over \thep}(\ln(\zeta\bar\zeta))^2]\cr}\eqn\sigto $$

$$\eqalign{ K &\approx 1  - 2 {(\thepc)\over (\thep)} \Re ( {d f_2\over 
du}) + 4 (\Re a_1)(\Re {d^2 f_2\over du^2}) \cr
&+ 2 {(1 - \zeta\bar\zeta)\over(\thep)^2} \Re({d^2 (f^{(2)})^2\over du^2}) +
{2 \zeta\bar\zeta\over(\thep)^2}\Re(f^{(2)}{d^2\overline{f^{(2)}}\over
du^2}) \cr
&+ 4 {(\thepc)\over (\thep)} \Re (i\overline{a^{(1)}} {d^3\overline{f^{(2)}}\over
du^3 }) \cr &+ 2 a^{(1)} \overline{a^{(1)}} \Re({d^3 f^{(2)}\over du^3})[2
\ln(\zeta\bar\zeta) + \half {\thepc\over
\thep}(\ln(\zeta\bar\zeta))^2]\cr}\eqn\kato$$
with 
$$a_1 = \sum_{j=1}^4 a^{(j)}; \qquad \qquad f_2 = \sum_{j=2}^4 f^{(j)}$$
In order to avoid the singularities deriving from the logarithmic terms,
we
can choose the first
term of the function $f_2$ as 
$ f^{(2)} = i k u$ with $k$ a real constant and leaving $f^{(3)}$ and $f^{(4)}$ 
as arbitrary functions of $u$.
Then, we have a regular fourth-order solution 
given by  
$$L \approx ({2 k^2\over\thep} + i k) {\bar\zeta\; u\over (\thep)^2}  
+ {a_1\over\zeta} + {\bar\zeta\over(\thep)^2} f_3(u) + {\thepc\over\thep}
{a_1\over\zeta} ({d f_3\over du } + i k)\eqn\elfin$$
$$\Phi \approx  - ik {a_1\over\zeta^2} 
      - {a_1\over\zeta^2}{d f_3\over du}
     + {a_1\over\zeta}({\bar\zeta\over\thep} - {1\over\zeta})\eqn\fifin$$
$$\Sigma \approx k [{(\thepc)\; u\over (\thep)} - 2 (\Re 
a^{(1)})]  + {(\thepc)\over 
(\thep)} \Im ( f_3(u)) - 2 (\Re a_1)(\Im {d f_3(u)\over du})\eqn\sigfi$$
$$K \approx - 4 {(1 - \zeta\bar\zeta) k^2\over(\thep)^2} 
+ 1 - 2 {(\thepc)\over (\thep)} \Re ( {d f_3\over 
du}) + 4 (\Re a_1)(\Re {d^2 f_3\over du^2})\eqn\kafin$$
and 
$$\Psi_4 \approx {\bar a_1\over r}\big({\thep\over\bar\zeta}\big)^2
{d^3 \overline{f_3}(u)\over du^3}\eqn\curfin$$
with 
$$a_1 = \sum_{j=1}^4 a^{(j)}; \qquad \qquad f_3 = f^{(3)} + f^{(4)}$$
At this point we have a non-flat twisting and regular solution of type N.
We have assumed that there are no non-zero invariants of the
curvature tensor and
used only the gauge invariants. 
Further calculations indicate that from the ansatz \ansa\ we have started
with
it is not possible to generate well behaved higher order corrections of
this solution. A new ansatz is needed in order to continue the study of
the twisting type N solutions using this iterative method.

We can only conclude that there is not yet
exact and regular twisting type N solutions. We would need to improved
our methods to seek for regular twisting type N solutions or 
to find a consistent
proof of the non-existence of regular type N solutions with non vanishing
twist.

NOTE ADDED: After this work was completed, we learned of the work by 
J. Bi\v c\'ak and V. Pravda [\BIPR], in which the scalar curvature
invariants for type N are studied.

\medskip
\noindent{\bf Acknowledgements}
\medskip
We would like to thank Daniel Finley and Maciej Przanowski for very
useful
conversations.
This work was partially suported by SNI and CONACyT-M\'exico.

\bigskip

\refout

\bye